\documentclass[12pt]{article}
\usepackage{amssymb}
\usepackage{epsfig}
\input epsf
\begin{document}
\title{
\hfill{}\\
\sc
A remark on the isotropic model
\vspace*{0.3cm}}
\author{ {\sc Alexander Moroz\thanks{www.amolf.nl/research/photonic\_materials\_theory/moroz/moroz.html}
\thanks{moroz@amolf.nl}}
\vspace*{0.3cm}}
\date{
\protect\normalsize
\it FOM Institute AMOLF, 
Kruislaan 407,  1098 SJ Amsterdam,  The Netherlands\\
and \\
I. Institut f\"{u}r Theoretische Physik,  Jungiusstrasse 9, 
Universit\"{a}t Hamburg, D-20355 Hamburg, Germany\thanks{present address}}

\maketitle

\begin{center}
{\large\sc abstract}
\end{center}
The applicability of the so-called isotropic complete photonic-band-gap 
(CPBG) models [S. John and J. Wang, Phys. Rev. Lett. {\bf 64}, 2418 (1990)]
 to capture essential features of the spontaneous emission (SE) of a fluorescent
atom or molecule near a  band-gap-edge of a CPBG structure is discussed.

\vspace*{0.6cm}

\noindent PACS number: 32.80.-t - Photon interactions with atoms \hfill\\
\noindent PACS number:  42.70.Qs - Photonic bandgap materials \hfill\\

\vspace*{1.9cm}

\thispagestyle{empty}
\baselineskip 20pt

\newpage

\noindent In their influential article \cite{JW}, John and Wang
considered  the quantum electrodynamics (QED) of an atom,
minimally coupled to the radiation filed, in the presence of a complete 
photonic-band-gap (CPBG).  In the latter case, there is a frequency interval in which,
independently of the photon direction and polarization, no photon modes can
propagate. Such a  QED vacuum differs significantly from
the conventional QED vacuum.  In  the vicinity of a  CPBG edge $\omega_c$ 
a number of exotic phenomena were predicted, among others, 
radical changes in  the spontaneous emission (SE) and an anomalous
Lamb shift.  (It is worthwhile to notice that some of the exotic
phenomena have been discussed much earlier in less known papers 
by Bykov \cite{Byk}.)  Atom properties were shown
to depend strongly on the exponent $\eta$ 
of the density of states (DOS) asymptotic  
\begin{equation}
\rho(\omega)\approx \mbox{const}\,  |\omega-\omega_c|^\eta
\label{jwf}
\end{equation}
near the CPBG edge. Calculations involving photonic crystals
in more than one dimension (1D) are notoriously difficult.
Therefore, in order to determine $\eta$,  John and Wang 
made use of approximations, subsequently employed in a number of recent 
discussions of the SE near a CPBG edge  \cite{JQ},
and often called isotropic and anisotropic CPBG models.
In the first model, the DOS near the band edge
$\omega_c$  is obtained from an approximated dispersion relation 
of the CPBG material 
$\omega_{\bf k} \approx \omega_c + A({\bf k}-{\bf k}_0)^2,$
where $A\approx \omega_c/{\bf k}_0^2$ and ${\bf k}_0$ is a vector at the
Brillouin zone boundary. The second model is then a slight generalization
of the first one. 

However, already in the Wigner-Weisskopf approximation
the SE of an isolated fluorescent atom (or molecule) in a fixed position 
${\bf r}$ within the unit cell  is determined    
by the local DOS (LDOS) and not the DOS  \cite{GL}. 
Considerations based on the DOS  can only be valid in the two hypothetical cases 
which are difficult to achieve:
(i) when the atom is allowed to freely propagate within a 
CPBG structure, as in experiments with 
cavity QED, and  (ii) when atoms are distributed
homogeneously within the entire unit cell. Only then the averaged 
and not the local properties of the QED vacuum within the unit cell
are probed.  The  fluorescent atoms are usually not distributed uniformly. 
If the atoms can be considered as  independent and are only radiatively
coupled to the crystal, the measured SE is determined by the weighted average 
(with the atomic position  probability distribution) of the
LDOS.  

In contrast to the DOS, the  LDOS behavior  near a  CPBG edge 
shows more complex behavior. Within a given frequency band,
the LDOS as a function of ${\bf r}$ exhibits  minima and maxima, their
number depending on the order of the band starting from the lowest one
\cite{AMe}. The LDOS behavior  is especially sensitive to 
${\bf r}$ in the vicinity of the LDOS minima, where the  LDOS approaches 
zero. If sufficiently close to a minimum of the LDOS, a shift in position by 
$10^{-4}$ of the unit cell length (i.e., a shift by one atom for optical photonic crystals)
can cause a change in the LDOS by  a factor of $3$ or higher.
Surprisingly enough, except at the exact position of a minimum,  the 
LDOS  near a band  edge has  asymptotic described by Eq. (\ref{jwf}) and the 
value of $\eta$   is still {\em universal}.  Therefore, at least in 1D, the predictions derived 
from the isotropic model hold true.

I should like to thank S. John and O. Toader for pointing to an error
in my original calculation.
This work is part of the research program by the Foundation for Fundamental 
Research on Matter which  was made possible by financial support from the 
Netherlands Organization for Scientific Research\footnote{Fortran code to calculate the LDOS 
asymptotic at band edges  is available upon request.}.

\end{document}